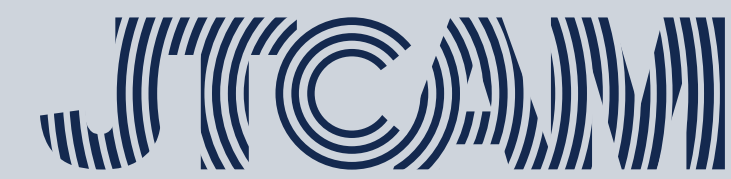

# A gradient-enhanced approach for stable finite element approximations of reaction-convection-diffusion problems

**Soheil Firooz[1], Batmanathan Daya Reddy[2], and Paul Steinmann[1,3]**

[1] Institute of Applied Mechanics, Friedrich-Alexander-Universität Erlangen–Nürnberg, Erlangen, Germany
[2] Department of Mathematics and Applied Mathematics and Centre for Research in Computational and Applied Mechanics, University of Cape Town, South Africa
[3] Glasgow Computational Engineering Center, James Watt School of Engineering, University of Glasgow, United Kingdom

Finite element solution of reaction–convection–diffusion equations are often unstable or inaccurate, particularly in regimes dominated by convection or reaction. This work addresses this challenge by developing a micromorphic-based finite element formulation that stabilizes the problem through gradient enhancement of the primary field via an auxiliary variable. Unlike many existing stabilization techniques that rely on problem-dependent parameters, the proposed approach offers a systematic and general framework. Theoretical analysis establishes the well-posedness of the coupled formulation and provides an error estimate, ensuring a sound mathematical basis. Numerical studies in one and two dimensions confirm that the method achieves high accuracy and enhanced stability across a wide range of conditions, including convection- and reaction-dominated cases. These findings suggest that micromorphic gradient enhancement can serve as a robust stabilization strategy, extending the applicability of finite element methods to more complex transport problems.

## 1 Introduction

*Reaction-convection-diffusion* equations play a central role in physical and mathematical modeling of transport and reaction phenomena across a wide range of disciplines from heat transfer and fluid flow to mass transport in environmental and chemical/biological systems. *Diffusion* commonly represents the spreading of species within the domain due to concentration gradients. *Convection* accounts for the transport of species driven by the movement of the fluid medium. *Reaction* refers to the chemical, biological or physical interactions through which species are produced or consumed during the process.

In reaction-convection-diffusion problems, if the diffusive effects are negligible compared to convective or reactive effects, the problem is referred to as convection-dominated or reaction-dominated, respectively (Christie et al. 1976; Heinrich et al. 1977; Johnson et al. 1984). A well-known challenge that arises in such problems is that the solutions develop sharp gradients and steep internal or boundary layers. In such scenarios, standard Galerkin finite element methods (FEM) fail to fully resolve the localized features leading to nonphysical, oscillatory solutions with spurious overshoots and undershoots (Gresho and Lee 1981; Brooks and Hughes 1982).

Over the past few decades, a variety of stabilization techniques have been developed to address the instabilities associated particularly with convection-diffusion equations. A classical remedy to eliminate such oscillations is to introduce artificial diffusion in the equations. However, excessive artificial diffusion can smear the solution which affects the solution accuracy.

One of the most well-established stabilization techniques to overcome instabilities in convection-dominated problems is the streamline-upwind Petrov–Galerkin (SUPG) introduced by Brooks and Hughes (1982); Hughes (1982). The SUPG method adds a perturbation term aligned with the flow/streamline direction to the test functions which effectively introduces a minimal targeted diffusion along streamlines. The resulting formulation maintains consistency and greatly reduces global oscillations for high Peclet number flows. In narrow regions such as sharp internal layers or boundary layers, the SUPG solution may still exhibit small overshoot/undershoot profiles because the stabilization is primarily in the flow direction and lacks dissipation in the crosswind direction (Burman 2010).

Another approach developed to address the numerical instabilities that arise in convection-dominated and reaction-dominated problems is the Galerkin/Least-Squares (GLS) method (Hughes et al. 1989; Pironneau et al. 1992). The GLS method enhances the standard Galerkin formulation via incorporating additional least-squares terms involving the residual of the governing equations (Franca and Dutra Do Carmo 1989). Unlike upwinding or artificial diffusion methods, the GLS method retains the variational consistency of the formulation. A critical and nontrivial aspect of the GLS method is the choice of the stabilization parameter which typically depends on local mesh size, the magnitude of convection and reaction coefficients, and sometimes solution features such as local gradients. Moreover, due to its global formulation which is advantageous for convergence analysis, GLS fails to offer fine local control over the stabilization effects. For such cases, other methods such as SUPG or the discontinuous Galerkin finite element method (DGFEM) (Johnson and Pitkäranta 1986; Cockburn 2003) may provide better adaptability and sharper resolution.

The so-called unusual stabilization method (Franca and Farhat 1995; Franca and Farhat 1994) maintains the original Galerkin framework but augments it with stabilizing terms that are not necessarily derived from a least-squares or residual-based interpretation. A key feature of this method is its variational consistency where the added stabilization terms vanish when the exact solution is substituted into the formulation thereby preserving convergence and error properties. Many studies have been carried out to improve the stability and accuracy of the unusual stabilization method, see (Franca and Valentin 2000; Duan et al. 2012; Hsieh and Yang 2016) among others.

Several other techniques have been developed to deal with instabilities associated with convection-dominated and reaction-dominated problems such as the variational multiscale (VMS) method (Hughes et al. 1998; John et al. 2006; Z.-J. Chen et al. 2018; Du et al. 2015), local projection stabilization (LPS) method (Knobloch and Lube 2009; Çıbık and Kaya 2011), the symmetric stabilization method (Burman and Fernández 2009; Guermond 1999; Codina 2000b), the residual bubble free (RFB) method (Franca et al. 1998; Brezzi et al. 1999), high-resolution Petrov–Galerkin (HRPG) method (Nadukandi et al. 2010; Nadukandi et al. 2012), weak Galerkin FEM (G. Chen et al. 2017), and variational subgrid scale method (Hauke and García-Olivares 2001). See (Codina 1998; Codina 2000a; John and Schmeyer 2008) for comparative studies on some stabilizing finite element methods. Despite these advances, challenges still remain in achieving stability without compromising accuracy. Some methods require tuning of problem-dependent parameters such as the SUPG weighting parameter or shock-capturing coefficients. Such limitations motivate the development of improved stabilization strategies that selectively target oscillations and internal layers without smearing the sharp gradients.

In a recent contribution (Firooz et al. 2025), a micromorphic-based artificial diffusion (MMAD) method has been proposed for stable and accurate finite element approximation of convection-diffusion problems. The method is, in principle, a gradient-enhanced extension of the mean-zero artificial diffusion method presented in (Firooz et al. 2024), and which is equivalent to a projected artificial diffusion approach. Motivated by the micromorphic approach (Forest 2009; Forest 2016; Forest and Sab 2020), the modification entails the introduction of a micromorphic-type variable and enhancement in terms of the variable and its gradient. A number of numerical examples demonstrate the excellent performance of the approach, in the convection-dominated range and for pure advection problems.

In this work the formulation in (Firooz et al. 2025) is extended to treat reaction-convection-diffusion problems. Of particular interest are situations in which reaction or convection dominate.

The formulation is shown to be well-posed and convergent for equal-order (linear) approximation of the original variable and that introduced for the enhancement. A number of numerical examples are presented, for situations of reaction- or advection- dominated flows. These illustrate the accurate and stable performance of the method, with oscillations or wiggles largely suppressed in regions of steep gradients or localized layers.

The structure of the rest of this work is as follows. Section 2 introduces the problem and presents the governing equations for reaction-convection-diffusion problems. The micromorphic-based artificial diffusion method together with details of the well-posedness and convergence of the method are the subject of Section 3. Section 4 illustrates the performance of the MMAD method through a set of numerical examples. Section 5 summarizes the the key contributions of the work.

## 2 Reaction-convection-diffusion problems

Let $\mathcal{B}$ be a bounded domain with the boundary $\partial\mathcal{B}$ and $\boldsymbol{n}$ be the outward unit normal to $\partial\mathcal{B}$. For a scalar quantity $\varphi$, the general *dimensionless* form of the steady-state reaction-convection-diffusion equation reads

$$\mathrm{Da}\varphi + \boldsymbol{u}\cdot\nabla\varphi - \tfrac{1}{\mathrm{Pe}}\Delta\varphi = F \qquad \text{in } \mathcal{B}, \tag{1a}$$

$$\varphi = \varphi_\mathrm{p} \qquad \text{on } \partial\mathcal{B}_\mathrm{D}, \tag{1b}$$

$$\nabla\varphi\cdot\boldsymbol{n} = \frac{\partial\varphi}{\partial\boldsymbol{n}} = t_\mathrm{p} \qquad \text{on } \partial\mathcal{B}_\mathrm{N}, \tag{1c}$$

where $\mathrm{Pe} = UL/D$ is the Peclet number, $\mathrm{Da} = BL/U$, the Damköhler number, $\boldsymbol{u} = \boldsymbol{U}/U$, the dimensionless velocity vector, $\boldsymbol{U}$ the divergence-free velocity, $D$ the diffusion coefficient, $B$ the reaction coefficient, $L$ a characteristic length of the spatial domain, $U$ a characteristic velocity, $\boldsymbol{n}$ the unit vector normal to the boundary $\partial\mathcal{B}$, $\varphi_\mathrm{p}$ the prescribed value of $\varphi$ and $t_\mathrm{p}$ the prescribed flux. $\partial\mathcal{B}_\mathrm{D}$ and $\partial\mathcal{B}_\mathrm{N}$ denote the Dirichlet and Neumann parts of the boundary $\partial\mathcal{B}$, respectively with $\partial\mathcal{B}_\mathrm{D}\cap\partial\mathcal{B}_\mathrm{N} = \emptyset$ and $\partial\mathcal{B}_\mathrm{D}\cup\partial\mathcal{B}_\mathrm{N} = \partial\mathcal{B}$. The first term in Equation (1) accounts for reaction, the second term accounts for convection, the third term accounts for diffusion and the term on the right is a source term. For consistency with the dimensionless formulation employed in our finite element method analysis–which proves convenient in the subsequent development–we define the dimensionless mesh size as $h = \overline{h}/L$, where $\overline{h}$ denotes the physical element size and $L$ the characteristic length of the domain. Note that the methodology presented in this paper is not restricted to steady-state problems. There are no inherent obstacles to applying the method to transient problems (see (Firooz et al. 2025) for the application of our approach to transient problems).

The Peclet number quantifies the relative influence of convection to diffusion whereas the Damköhler number quantifies the relative influence of reaction to convection. If $\mathrm{Pe}\gg 1$ and $\mathrm{Da}\ll 1$, convection dominates over diffusion and reaction; sharp gradients or boundary layers are likely to emerge in the solution. This is common in high-speed flows or large-scale systems. If $\mathrm{Da}\gg 1$ and $\mathrm{Pe}\ll 1$, reaction dominates over convection and diffusion; reaction kinetics are fast compared to transport processes and the species may be consumed near boundaries or in localized regions before being transported further, which also leads to sharp concentration gradients and boundary layers. This is common in catalytic reactors, biological systems or porous media transport systems. If $\mathrm{Da}\gg 1$ and $\mathrm{Pe}\gg 1$, the system is advection- and reaction-dominated; the solution often exhibits steep concentration gradients along streamlines and thin reactive boundary layers. This regime is commonly encountered in high-speed combustion processes, industrial chemical reactors or chemical vapor deposition systems. If $\mathrm{Da}\ll 1$ and $\mathrm{Pe}\ll 1$, the system is diffusion dominated; the solution exhibits uniform concentration distributions and smooth profiles where sharp gradients or boundary layers are generally absent. This regime is typical in microfluidic devices, low-speed flows or dilute biological systems.

To calculate the weak form of Equation (1), we multiply Equation (1a) with a test function $\delta\varphi$ and integrate it over $\mathcal{B}$ which, using the divergence theorem and Neumann boundary condition,

reads

$$\int_{\mathcal{B}} \mathrm{Da}\varphi\delta\varphi \,\mathrm{d}v + \int_{\mathcal{B}} \boldsymbol{u}\cdot\nabla\varphi\delta\varphi \,\mathrm{d}v + \tfrac{1}{\mathrm{Pe}}\int_{\mathcal{B}} \nabla\varphi\cdot\nabla\delta\varphi \,\mathrm{d}v = \int_{\mathcal{B}} F\delta\varphi \,\mathrm{d}v + \int_{\partial\mathcal{B}_{\mathrm{N}}} t_{\mathrm{p}}\delta\varphi \,\mathrm{d}a, \tag{2}$$

for $\varphi = \varphi_{\mathrm{p}}$ on $\partial\mathcal{B}_{\mathrm{D}}$ where the test function $\delta\varphi$ satisfies $\delta\varphi = 0$ on $\partial\mathcal{B}_{\mathrm{D}}$. Here $t_p$ is the prescribed boundary flux; see Equation (1c).

## 3 The micromorphic-based artificial diffusion method

The MicroMorphic-based Artificial Diffusion method (MMAD) formulation for reaction-convection-diffusion problems follows closely that for convection-diffusion problems, as set out in (Firooz et al. 2025), but for convenience is presented in self-contained fashion. We assume in this development a homogeneous Dirichlet boundary condition on all of $\partial\mathcal{B}$. The extension to more general boundary conditions is straightforward.

We start by defining the micromorphic-type variable $\boldsymbol{g}$ and the generalized strain-like variable $\boldsymbol{e} = \nabla\varphi - \boldsymbol{g}$. Then for the special case of a reaction-diffusion problem, for which a minimization problem exists, the total potential is given by

$$\Psi = \Psi_0 + \Psi_{\mathrm{MM}} \tag{3}$$

where

$$\begin{cases} \Psi_0 = \dfrac{1}{2\mathrm{Pe}}|\nabla\varphi|^2 + \dfrac{\mathrm{Da}}{2}|\varphi|^2 + F\varphi, \\ \Psi_{\mathrm{MM}} = \dfrac{1}{2}[\nabla\varphi - \boldsymbol{g}]\cdot\boldsymbol{H}\cdot[\nabla\varphi - \boldsymbol{g}] + \dfrac{1}{2}\boldsymbol{g}\cdot\boldsymbol{K}\cdot\boldsymbol{g} + \dfrac{1}{2}\nabla\boldsymbol{g} : \mathbb{A} : \nabla\boldsymbol{g} \end{cases} \tag{4}$$

with $\boldsymbol{H}$, the micromorphic-type (second-order) coupling tensor, $\boldsymbol{K}$, the micromorphic-type (second-order) tensor and $\mathbb{A}$, the micromorphic-type (fourth-order) "stiffness" tensor. We define $\boldsymbol{H}$, $\boldsymbol{K}$ and $\mathbb{A}$ such that they are symmetric and positive-definite; that is, for any vector $\boldsymbol{a}$ and second-order tensor $\boldsymbol{A}$ we have

$$[\boldsymbol{H}\cdot\boldsymbol{a}]\cdot\boldsymbol{a} \geqslant H_0|\boldsymbol{a}|^2, \quad [\boldsymbol{K}\cdot\boldsymbol{a}]\cdot\boldsymbol{a} \geqslant K_0|\boldsymbol{a}|^2, \quad [\mathbb{A}:\boldsymbol{A}]:\boldsymbol{A} \geqslant A_0|\boldsymbol{A}|^2, \tag{5}$$

where $H_0$, $K_0$ and $A_0$ are positive constants. The micromorphic-type coupling tensor $\boldsymbol{H}$ enforces compatibility between the primary field and the micromorphic variable. In the discrete formulation, $\boldsymbol{H}$ is selected such that it incorporates the effects of upwinding and reaction terms, ensuring appropriate stabilization. On the other hand, the micromorphic-type "stiffness" tensor $\mathbb{A}$ governs the spatial scale of micromorphic effects. Moreover, through appearing in the gradient terms, it also controls the degree to which the micromorphic effects influence the solution. A key feature of our proposed MMAD method is that the micromorphic effects can be easily controlled by choosing appropriate values for $\boldsymbol{H}$, $\boldsymbol{K}$ and $\mathbb{A}$ thereby ensuring stable and accurate solutions suited to the problem at hand.

**Remark 1** The choice $\mathbb{A} = \mathbb{O}$, $\boldsymbol{K} = \boldsymbol{0}$ and $\boldsymbol{H} = p\boldsymbol{I}$ reduces the MMAD formulation to the MZAD approach for pure advection coupled-problems introduced in (Firooz et al. 2024).

The first variation of the total energy

$$\Pi(\varphi, \boldsymbol{g}) = \int_{\mathcal{B}} [\Psi_0(\varphi) + \Psi_{\mathrm{MM}}(\varphi, \boldsymbol{g})] \,\mathrm{d}v, \tag{6}$$

leads to a weak formulation. With the addition of the convective term this gives the full MMAD weak formulation

$$\int_{\mathcal{B}} \mathrm{Da}\varphi\delta\varphi \,\mathrm{d}v + \int_{\mathcal{B}} \boldsymbol{u}\cdot\nabla\varphi\delta\varphi \,\mathrm{d}v + \int_{\mathcal{B}} \tfrac{1}{\mathrm{Pe}}\nabla\varphi\cdot\nabla\delta\varphi \,\mathrm{d}v + \int_{\mathcal{B}} [\boldsymbol{H}\cdot[\nabla\varphi - \boldsymbol{g}]]\cdot\nabla\delta\varphi \,\mathrm{d}v$$

$$= \int_{\mathcal{B}} F\delta\varphi \,\mathrm{d}v, \tag{7a}$$

$$\int_{\mathcal{B}} [-\boldsymbol{H}\cdot[\nabla\varphi - \boldsymbol{g}] + \boldsymbol{K}\cdot\boldsymbol{g}]\cdot\delta\boldsymbol{g} \,\mathrm{d}v + \int_{\mathcal{B}} [\mathbb{A}:\nabla\boldsymbol{g}] : \nabla\delta\boldsymbol{g} \,\mathrm{d}v = 0. \tag{7b}$$

We define the spaces $\Phi = \mathcal{H}_0^1(\mathcal{B}) = \{\varphi \in L^2(\mathcal{B}) \,|\, \partial_{x_i}\varphi \in L^2(\mathcal{B}),\ \varphi = 0 \text{ on } \partial\mathcal{B}\}$ and $G = \{\boldsymbol{g} \,|\, g_i \in \mathcal{H}^1(\mathcal{B})\}$ as well as the norms $\|\varphi\|_\Phi = \|\nabla\varphi\|_{L^2} = (\int_\mathcal{B} |\nabla\varphi|^2 \,\mathrm{d}v)^{1/2}$, $\|\boldsymbol{g}\|^2_{L^2} = \int_\mathcal{B} |\boldsymbol{g}|^2 \,\mathrm{d}v$, and $\|\boldsymbol{g}\|^2_G = \|\boldsymbol{g}\|^2_{L^2} + \|\nabla\boldsymbol{g}\|^2_{L^2}$. Our objective is to seek $\varphi \in \Phi$ and $\boldsymbol{g} \in G$ that satisfy Equation (7) for all $\delta\varphi \in \Phi$ and $\delta\boldsymbol{g} \in G$.

## 3.1 Discrete MMAD formulation

We denote the discrete conforming approximations of $\varphi$ and $\boldsymbol{g}$ as $\varphi_h$ and $\boldsymbol{g}_h$, respectively, together with their corresponding test functions $\delta\varphi_h$ and $\delta\boldsymbol{g}_h$. For the discretized MMAD formulation, the objective is: for all $\delta\varphi_h \in \Phi^h$ and $\delta\boldsymbol{g}_h \in G^h$, find $\varphi_h \in \Phi^h \subset \Phi$ and $\boldsymbol{g}_h \in G^h \subset G$ such that

$$
\begin{aligned}
&\int_\mathcal{B} \mathrm{Da}\varphi_h\delta\varphi_h \,\mathrm{d}v + \int_\mathcal{B} \boldsymbol{u}\cdot\nabla\varphi_h\delta\varphi_h \,\mathrm{d}v + \int_\mathcal{B} \tfrac{1}{\mathrm{Pe}}\nabla\varphi_h\cdot\nabla\delta\varphi_h \,\mathrm{d}v + \int_\mathcal{B} [\boldsymbol{H}\cdot[\nabla\varphi_h - \boldsymbol{g}_h]]\cdot\nabla\delta\varphi_h \,\mathrm{d}v \\
&\qquad = \int_\mathcal{B} F\delta\varphi_h \,\mathrm{d}v, && \text{(8a)}\\
&\int_\mathcal{B} [-\boldsymbol{H}\cdot[\nabla\varphi_h - \boldsymbol{g}_h] + \boldsymbol{K}\cdot\boldsymbol{g}_h]\cdot\delta\boldsymbol{g}_h \,\mathrm{d}v + \int_\mathcal{B} [\mathbb{A} : \nabla\boldsymbol{g}_h] : \nabla\delta\boldsymbol{g}_h \,\mathrm{d}v = 0. && \text{(8b)}
\end{aligned}
$$

### 3.1.1 Well-posedness and convergence analysis

To elaborate on the well-posedness of the formulation, we define the space $\widehat{\Phi} = \Phi \times G$. For $\widehat{\boldsymbol{\varphi}} := (\varphi, \boldsymbol{g}) \in \widehat{\Phi}$ the norm $\|\bullet\|_{\widehat{\Phi}}$ is defined by $\|\widehat{\boldsymbol{\varphi}}\|^2_{\widehat{\Phi}} = \|\varphi\|^2_\Phi + \|\boldsymbol{g}\|^2_G$. Then the fully discrete version of Equation (8) can be written as follows: for all $\widehat{\boldsymbol{\varphi}}_h \in \widehat{\Phi}^h$, find $\widehat{\boldsymbol{\varphi}}_h \in \widehat{\Phi}^h$ such that

$$
B\left(\widehat{\boldsymbol{\varphi}}_h, \delta\widehat{\boldsymbol{\varphi}}_h\right) = \ell(\delta\widehat{\boldsymbol{\varphi}}_h) \tag{9}
$$

where

$$
\begin{aligned}
B\left(\widehat{\boldsymbol{\varphi}}_h, \delta\widehat{\boldsymbol{\varphi}}_h\right) &= \int_\mathcal{B} \Big[\mathrm{Da}\varphi_h\delta\varphi_h + \boldsymbol{u}\cdot\nabla\varphi_h\delta\varphi_h + \tfrac{1}{\mathrm{Pe}}\nabla\varphi_h\cdot\nabla\delta\varphi_h + [\boldsymbol{H}\cdot[\nabla\varphi_h - \boldsymbol{g}_h]]\cdot[\nabla\delta\varphi_h - \delta\boldsymbol{g}_h] \\
&\qquad + [\boldsymbol{K}\cdot\boldsymbol{g}_h]\cdot\delta\boldsymbol{g}_h + [\mathbb{A} : \nabla\boldsymbol{g}_h] : \nabla\delta\boldsymbol{g}_h\Big] \,\mathrm{d}v, && \text{(10a)}\\
\ell(\delta\widehat{\boldsymbol{\varphi}}_h) &= \int_\mathcal{B} F\delta\varphi_h \,\mathrm{d}v. && \text{(10b)}
\end{aligned}
$$

To show well-posedness, one needs to show that $B(\bullet,\bullet)$ is coercive and continuous, and $\ell(\bullet)$ is continuous. That is,

$$
\begin{aligned}
&B(\varphi_h, \boldsymbol{g}_h; \varphi_h, \boldsymbol{g}_h) \geqslant M\, \|\widehat{\boldsymbol{\varphi}}_h\|^2_{\widehat{\Phi}}, && \text{(11a)}\\
&B(\varphi_h, \boldsymbol{g}_h; \delta\varphi_h, \delta\boldsymbol{g}_h) \leqslant m\, \|\widehat{\boldsymbol{\varphi}}_h\|_{\widehat{\Phi}}\, \|\delta\widehat{\boldsymbol{\varphi}}_h\|_{\widehat{\Phi}}, && \text{(11b)}\\
&\ell(\varphi_h, \boldsymbol{g}_h) \leqslant c\, \|\widehat{\boldsymbol{\varphi}}_h\|_{\widehat{\Phi}}, && \text{(11c)}
\end{aligned}
$$

with $M$, $m$ and $c$ being positive constants. For the coercivity of $B(\bullet,\bullet)$, we start with

$$
\begin{aligned}
B(\widehat{\boldsymbol{\varphi}}_h, \widehat{\boldsymbol{\varphi}}_h) &= \int_\mathcal{B} \Big[\mathrm{Da}|\varphi_h|^2 + [\boldsymbol{u}\cdot\nabla\varphi_h]\varphi_h + \tfrac{1}{\mathrm{Pe}}|\nabla\varphi_h|^2 + [\boldsymbol{H}\cdot[\nabla\varphi_h - \boldsymbol{g}_h]]\cdot[\nabla\varphi_h - \boldsymbol{g}_h] \\
&\qquad + [\boldsymbol{K}\cdot\boldsymbol{g}_h]\cdot\boldsymbol{g}_h + [\mathbb{A} : \nabla\boldsymbol{g}_h] : \nabla\boldsymbol{g}_h\Big] \,\mathrm{d}v. && \text{(12)}
\end{aligned}
$$

We show that the second term on the right-hand side is zero, as follows. using the divergence theorem

$$
\begin{aligned}
\int_\mathcal{B} [\boldsymbol{u}\cdot\nabla\varphi]\varphi \,\mathrm{d}v &= \int_{\partial\mathcal{B}} \varphi^2[\boldsymbol{u}\cdot\boldsymbol{n}] \,\mathrm{d}a - \int_\mathcal{B} \varphi\nabla\cdot(\varphi\boldsymbol{u}) \,\mathrm{d}v \\
&= \int_{\partial\mathcal{B}} \varphi^2[\boldsymbol{u}\cdot\boldsymbol{n}] \,\mathrm{d}a - \int_\mathcal{B} \varphi^2\nabla\cdot\boldsymbol{u} \,\mathrm{d}v - \int_\mathcal{B} [\boldsymbol{u}\cdot\nabla\varphi]\varphi \,\mathrm{d}v. && \text{(13)}
\end{aligned}
$$

The first two terms on the right-hand side are zero, given respectively the homogeneous Dirichlet boundary condition on $\varphi$, and the incompressibility condition. Rearrangement gives the desired result. From Equation (12) and using Equation (13) we have

$$B(\widehat{\boldsymbol{\varphi}}_h, \widehat{\boldsymbol{\varphi}}_h) \geqslant \int_{\mathcal{B}} [\mathrm{Da}|\varphi_h|^2 + \tfrac{1}{\mathrm{Pe}}|\nabla\varphi_h|^2 + H_0|\nabla\varphi_h - \boldsymbol{g}_h|^2 + K_0|\boldsymbol{g}_h|^2 + A_0|\nabla\boldsymbol{g}_h|^2]\,\mathrm{d}v. \tag{14}$$

Using the inequality $2ab \leqslant \varepsilon a^2 + b^2/\varepsilon$ which implies $[a-b]^2 \geqslant [1-\varepsilon]a^2 + [1-\varepsilon^{-1}]b^2$ for any $\varepsilon$, we have $|\nabla\varphi_h - \boldsymbol{g}_h|^2 \geqslant [1-\varepsilon]|\nabla\varphi_h|^2 + [1-\varepsilon^{-1}]|\boldsymbol{g}_h|^2$. Consequently, Equation (14) becomes

$$\begin{aligned}
B(\widehat{\boldsymbol{\varphi}}_h, \widehat{\boldsymbol{\varphi}}_h) &\geqslant \int_{\mathcal{B}} \left[\mathrm{Da}|\varphi_h|^2 + [\tfrac{1}{\mathrm{Pe}} + H_0[1-\varepsilon]]|\nabla\varphi_h|^2 + [H_0[1-\varepsilon^{-1}] + K_0]|\boldsymbol{g}_h|^2 + A_0|\nabla\boldsymbol{g}_h|^2\right]\mathrm{d}v \\
&\geqslant [\tfrac{1}{\mathrm{Pe}} + H_0[1-\varepsilon]]||\varphi_h||^2_{\Phi} + [H_0[1-\varepsilon^{-1}] + K_0]||\boldsymbol{g}_h||^2_{L^2} + A_0||\nabla\boldsymbol{g}_h||^2_{L^2} \\
&\geqslant [\tfrac{1}{\mathrm{Pe}} + H_0[1-\varepsilon]]||\varphi_h||^2_{\Phi} + \min(H_0[1-\varepsilon^{-1}] + K_0, A_0)||\boldsymbol{g}_h||^2_{G},
\end{aligned} \tag{15}$$

choosing $\varepsilon$ such that $0 < \varepsilon < 1$, and $\boldsymbol{K}$ such that $K_0 + H_0[1-\varepsilon^{-1}] > 0$. Then

$$B(\widehat{\boldsymbol{\varphi}}_h, \widehat{\boldsymbol{\varphi}}_h) \geqslant M||\widehat{\boldsymbol{\varphi}}||^2_{\widehat{\Phi}}, \tag{16}$$

with $M = \min\left(1/\mathrm{Pe} + H_0[1-\varepsilon], \min\left(K_0 + H_0[1-\varepsilon^{-1}], A_0\right)\right)$. For continuity, we have

$$\begin{aligned}
B(\widehat{\boldsymbol{\varphi}}_h, \delta\widehat{\boldsymbol{\varphi}}_h) = \int_{\mathcal{B}} &[\mathrm{Da}\varphi_h\delta\varphi_h + \boldsymbol{u}\cdot\nabla\varphi_h\delta\varphi_h + \tfrac{1}{\mathrm{Pe}}\nabla\varphi_h\cdot\nabla\delta\varphi_h + [\boldsymbol{H}\cdot[\nabla\varphi_h - \boldsymbol{g}_h]]\cdot[\nabla\delta\varphi_h - \delta\boldsymbol{g}_h] \\
&+ [\boldsymbol{K}\cdot\boldsymbol{g}_h]\cdot\delta\boldsymbol{g}_h + [\mathbb{A} : \nabla\boldsymbol{g}_h] : \nabla\delta\boldsymbol{g}_h]\,\mathrm{d}v,
\end{aligned} \tag{17}$$

Then using the Cauchy–Schwarz inequality, we have

$$\begin{aligned}
|B(\widehat{\boldsymbol{\varphi}}_h, \delta\widehat{\boldsymbol{\varphi}}_h)| \leqslant \int_{\mathcal{B}} &[\mathrm{Da}|\varphi_h||\delta\varphi_h| + u_{\max}|\nabla\varphi_h||\delta\varphi_h| + \tfrac{1}{\mathrm{Pe}}|\nabla\varphi_h||\nabla\delta\varphi_h| \\
&+ H_{\max}|\nabla\varphi_h - \boldsymbol{g}_h||\nabla\delta\varphi_h - \delta\boldsymbol{g}_h| + K_{\max}|\boldsymbol{g}_h||\delta\boldsymbol{g}_h| + A_{\max}|\nabla\boldsymbol{g}_h||\nabla\delta\boldsymbol{g}_h|]\,\mathrm{d}v \\
\leqslant\ &\mathrm{Da}||\varphi_h||_{L^2}||\delta\varphi_h||_{L^2} + [u_{\max}||\delta\varphi_h||_{L^2} + \tfrac{1}{\mathrm{Pe}}||\nabla\delta\varphi_h||_{L^2}]||\nabla\varphi_h||_{L^2} \\
&+ 2H_{\max}||\widehat{\boldsymbol{\varphi}}_h||_{\widehat{\Phi}}||\delta\widehat{\boldsymbol{\varphi}}_h||_{\widehat{\Phi}} + K_{\max}||\boldsymbol{g}_h||_{L^2}||\delta\boldsymbol{g}_h||_{L^2} + A_{\max}||\nabla\boldsymbol{g}_h||_{L^2}||\nabla\delta\boldsymbol{g}_h||_{L^2} \\
\leqslant\ &[\mathrm{Da} + u_{\max} + \tfrac{1}{\mathrm{Pe}} + 2H_{\max} + K_{\max} + A_{\max}]||\widehat{\boldsymbol{\varphi}}_h||_{\widehat{\Phi}}||\delta\widehat{\boldsymbol{\varphi}}_h||_{\widehat{\Phi}},
\end{aligned} \tag{18}$$

with $u_{\max} = \max_i |u_i|$, $H_{\max} = \max_{ij} |H_{ij}|$, $K_{\max} = \max_{ij} |K_{ij}|$ and $A_{\max} = \max_{ijkl} |A_{ijkl}|$. Therefore $B(\bullet, \bullet)$ is continuous with constant

$$m = \mathrm{Da} + u_{\max} + \tfrac{1}{\mathrm{Pe}} + 2H_{\max} + K_{\max} + A_{\max}. \tag{19}$$

To show continuity of $\ell(\bullet)$, from Equation (10) we have, using the Cauchy–Schwarz inequality,

$$|\ell(\varphi_h)| = \left|\int_{\mathcal{B}} F\varphi_h\,\mathrm{d}v\right| \leqslant ||F||_{L^2}||\varphi_h||_{L^2}, \tag{20}$$

so that Equation (11c) is verified with $c = ||F||_{L^2}$.

### 3.1.2 Error analysis

The continuity and coercivity of $B(\bullet, \bullet)$ imply the standard finite element error estimate for $V^h, G^h$ comprising continuous piecewise-linear polynomial is (Reddy 1998)

$$||\widehat{\boldsymbol{\varphi}} - \widehat{\boldsymbol{\varphi}}_h||_{\widehat{\Phi}} \leqslant Ch. \tag{21}$$

Thus, we have convergence at a linear rate and the constant $C$ is given by $C = c[m/M]$, where $m$ and $M$ are the continuity and coercivity constants, respectively, and $c$ a constant that depends on the $\mathcal{H}^2$ semi-norm of $\widehat{\boldsymbol{\varphi}}$. Next, we consider the modeling error. The original problem is to find $\varphi_0 \in \Phi$ such that

$$a\left(\varphi_0, \delta\varphi\right) = \ell\left(\delta\varphi\right) \tag{22}$$

with

$$a(\varphi_0, \delta\varphi) = \int_{\mathcal{B}} [\mathrm{Da}\varphi_0\delta\varphi + [\boldsymbol{u}\cdot\nabla\varphi_0]\delta\varphi + \tfrac{1}{\mathrm{Pe}}\nabla\varphi_0\cdot\nabla\delta\varphi]\,\mathrm{d}v \quad \text{and} \quad \ell(\delta\varphi) = \int_{\mathcal{B}} F\delta\varphi\,\mathrm{d}v.$$

On the other hand, the MMAD problem is given by (7). Thus we have an error $[\varphi - \varphi_0]$ between the original (or actual) and MMAD solutions. We now estimate this error. Setting

$$a(\varphi, \delta\varphi) + b(\varphi, \boldsymbol{g}; \delta\varphi) = \ell(\delta\varphi), \quad \text{with} \quad b(\varphi, \boldsymbol{g}; \delta\varphi) = \int_{\mathcal{B}} [\boldsymbol{H}\cdot[\nabla\varphi - \boldsymbol{g}]]\cdot\nabla\delta\varphi\,\mathrm{d}v, \tag{23}$$

from Equations (22) and (23) we have

$$a(\varphi - \varphi_0, \delta\varphi) + b(\widehat{\boldsymbol{\varphi}}, \delta\varphi) = 0. \tag{24}$$

Now $a$ is coercive since

$$a(\varphi, \varphi) \geqslant M_0||\varphi||^2_{\Phi} \tag{25}$$

with $M_0 = 1/\mathrm{Pe}$ using Equation (13). Next, set $\delta\varphi = \varphi - \varphi_0$ in Equation (24); this gives

$$a(\varphi - \varphi_0, \varphi - \varphi_0) + b(\widehat{\boldsymbol{\varphi}}, \varphi - \varphi_0) = 0. \tag{26}$$

Thus, from Equations (25) and (26),

$$M_0\,||\varphi - \varphi_0||^2_{\Phi} \leqslant b(\widehat{\boldsymbol{\varphi}}, \varphi_0 - \varphi) \leqslant \beta_0\,||\widehat{\boldsymbol{\varphi}}||_{\widehat{\Phi}}\,||\varphi - \varphi_0||_{\Phi}, \tag{27}$$

where $\beta_0 = H_{\max}$. Hence,

$$||\varphi - \varphi_0||_{\Phi} \leqslant \frac{\beta_0}{M_0}||\widehat{\boldsymbol{\varphi}}||_{\widehat{\Phi}}. \tag{28}$$

Finally, combining Equations (21) and (28) we have, for the error between the original solution and its MMAD finite element approximation,

$$||\varphi_0 - \varphi_h||_{\Phi} \leqslant ||\varphi_0 - \varphi||_{\Phi} + ||\varphi - \varphi_h||_{\Phi} \leqslant \frac{\beta_0}{M_0}||\widehat{\boldsymbol{\varphi}}||_{\widehat{\Phi}} + Ch. \tag{29}$$

**Remark 2** The modeling error depends on $\beta_0/M_0 = H_{\max}\mathrm{Pe}$. Thus it can be controlled in the limit of small diffusivity ($\mathrm{Pe} \gg 1$) through an appropriate choice for $\boldsymbol{H}$. The finite element approximation error depends on $m/M$, where these constants are given in Equations (16) and (19). The choices for $\boldsymbol{H}$, $\boldsymbol{K}$ and $\mathbb{A}$ will be discussed in Section 3.2.

**Remark 3** We note that we have stability in the limit of vanishing reaction ($\mathrm{Da} \to 0$).

## 3.2 Choice of $\boldsymbol{H}$, $\boldsymbol{K}$ and $\mathbb{A}$

The next step to complete our approach is to elaborate on the choice of $\boldsymbol{H}$, $\boldsymbol{K}$ and $\mathbb{A}$. Following the works (Brooks and Hughes 1982; Tezduyar and Park 1986), we set $\boldsymbol{H} = \overline{k}_{\mathrm{c}}\widehat{\boldsymbol{u}}\otimes\widehat{\boldsymbol{u}} + \overline{k}_{\mathrm{r}}\boldsymbol{I}$, $\boldsymbol{K} = \boldsymbol{I}$, and $\mathbb{A} = \mathbb{I}$, where $\widehat{\boldsymbol{u}} = \boldsymbol{u}/|\boldsymbol{u}|$ and $\boldsymbol{I}$ and $\mathbb{I}$ being the second- and fourth-order identity tensors, respectively. The constants $\overline{k}_{\mathrm{c}}$ and $\overline{k}_{\mathrm{r}}$ read

$$\overline{k}_{\mathrm{c}} = \sum_{i=1}^{\mathrm{PD}} u_i h_i \gamma_i/2 \quad \text{and} \quad \overline{k}_{\mathrm{r}} = \frac{1}{\mathrm{Pe}}\Big[\frac{2}{3}\beta^2 + \frac{\beta^2}{\sinh^2\beta} - 1\Big] \tag{30}$$

with $\gamma_i = \coth(\alpha_i) - 1/\alpha_i$, $\alpha_i = \mathrm{Pe}h_i/2$, $u_i = \boldsymbol{e}_i^T\cdot\boldsymbol{u}$, and $\beta = \sqrt{(Bh^2)/(4D)}$. The subscripts "c" and "r" refer to convection and reaction, respectively, PD is the problem dimension, $\boldsymbol{e}_i$, the unit vectors in the finite element natural coordinate system and $h_i$, the dimensionless element characteristic length in each natural coordinate direction. Also, $D$ denotes the diffusion coefficient (in units of $\mathrm{m}^2/\mathrm{s}$), $B$ is the reaction coefficient (in units of $\mathrm{s}^{-1}$), and $h$ represents the dimensionless mesh size.

**Remark 4** Specifically, $h_i$ refers to the dimensionless characteristic element size in the $i$-th coordinate direction. Since we employ structured meshes in all numerical examples, the mesh is uniform in all directions, and thus $h_i = h$ for all $i$. In the case of unstructured meshes, defining an appropriate scalar representative mesh size becomes more involved.

# 4 Numerical results

In this section we evaluate the performance of the MMAD formulation through a set of numerical examples in one- and two-dimensional settings. Six different examples are examined. Figure 1 shows the boundary conditions and problem specifications associated with each example.

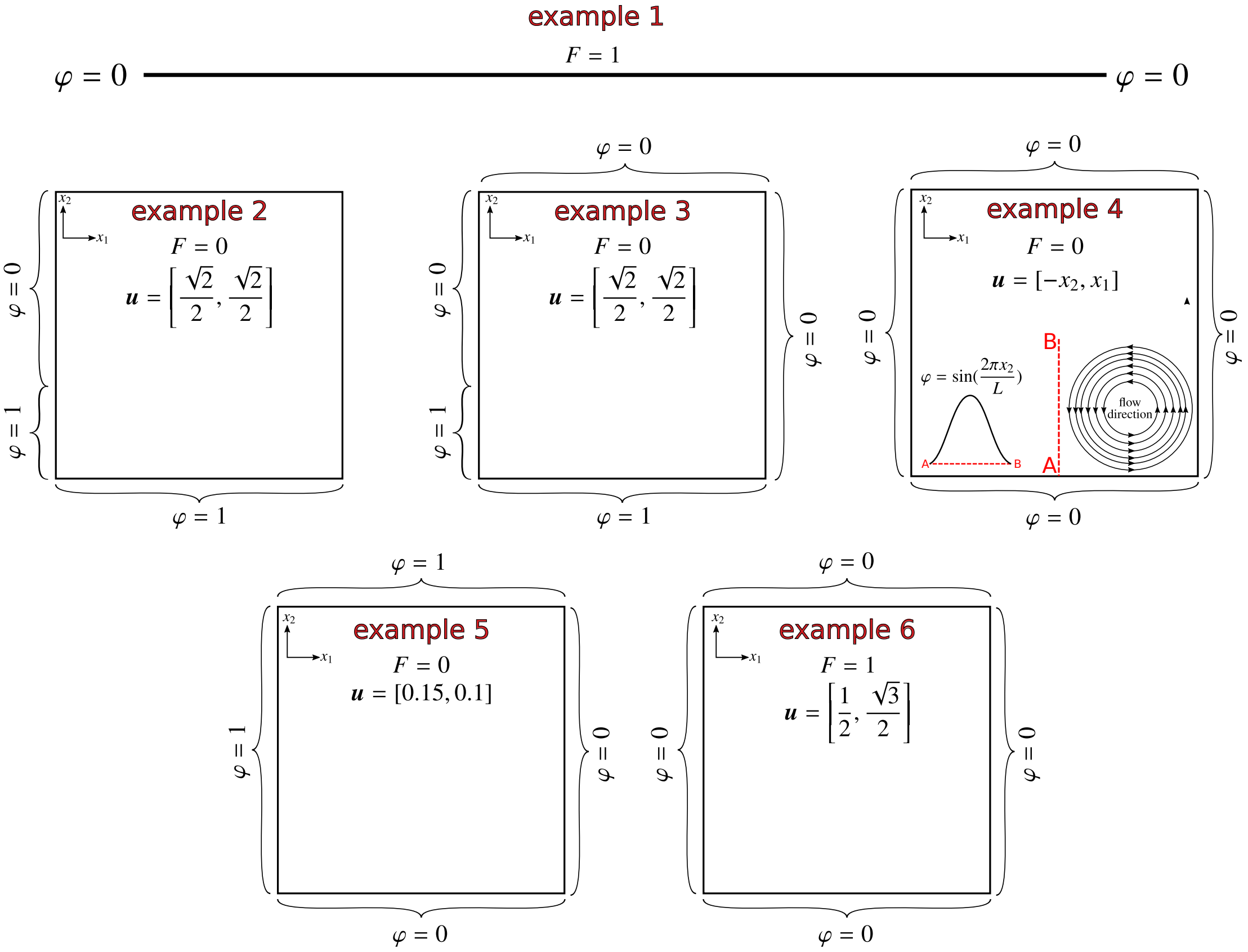

**Figure 1** Six different case studies for one-dimensional and two-dimensional reaction-convection-diffusion equations. The boundary conditions, the flow direction and the source term are specified for each case. For the 1D example, the domain spans $\mathcal{B} := [0, 1]$ and for 2D examples, the domain spans $\mathcal{B} := [0, 1] \times [0, 1]$.

The first example is 1D and the rest of the examples are 2D. In order to effectively examine the influence of reaction or convection dominance in the problem, different cases with various Damköhler numbers and Peclet numbers are considered. All simulations are carried out using our in-house finite element code. For the 1D example, the domain is discretized into 100 linear elements. For the 2D examples, the domain is discretized into a structured grid consisting of 1600 bi-linear quadrilateral elements. Throughout the examples, continuous piecewise-linear functions are employed for the finite element approximation of all fields.

**Example 1** We consider a 1D reaction-convection-diffusion problem. As shown in Figure 1, both ends of the domain are subject to Dirichlet boundary conditions $\varphi = 0$ and the source term is $F = 1$.

Figure 2 shows the solutions obtained by standard Galerkin FEM and the MMAD method for different values of Damköhler and Peclet numbers. The solid black line in each plot shows the exact solution and the dashed line with points shows numerical solutions. Five different cases are considered; two extreme cases in the top row and the bottom row representing a highly reaction-dominated regime and a highly convection-dominated regime, respectively; and three intermediate cases with comparable convection and reaction contributions. For all these cases the diffusion is very small. It is observed that the FEM solution shows small oscillations in the vicinity of the sharp gradients for the reaction-dominated case.

For the three intermediate cases, increasing the Peclet number damps out the oscillations on the left side but exacerbates the oscillations on the right side.

For the convection-dominated case, the FEM solution is completely oscillatory throughout the entire domain. The MMAD method for all cases gives stable, accurate and oscillation-free solutions.

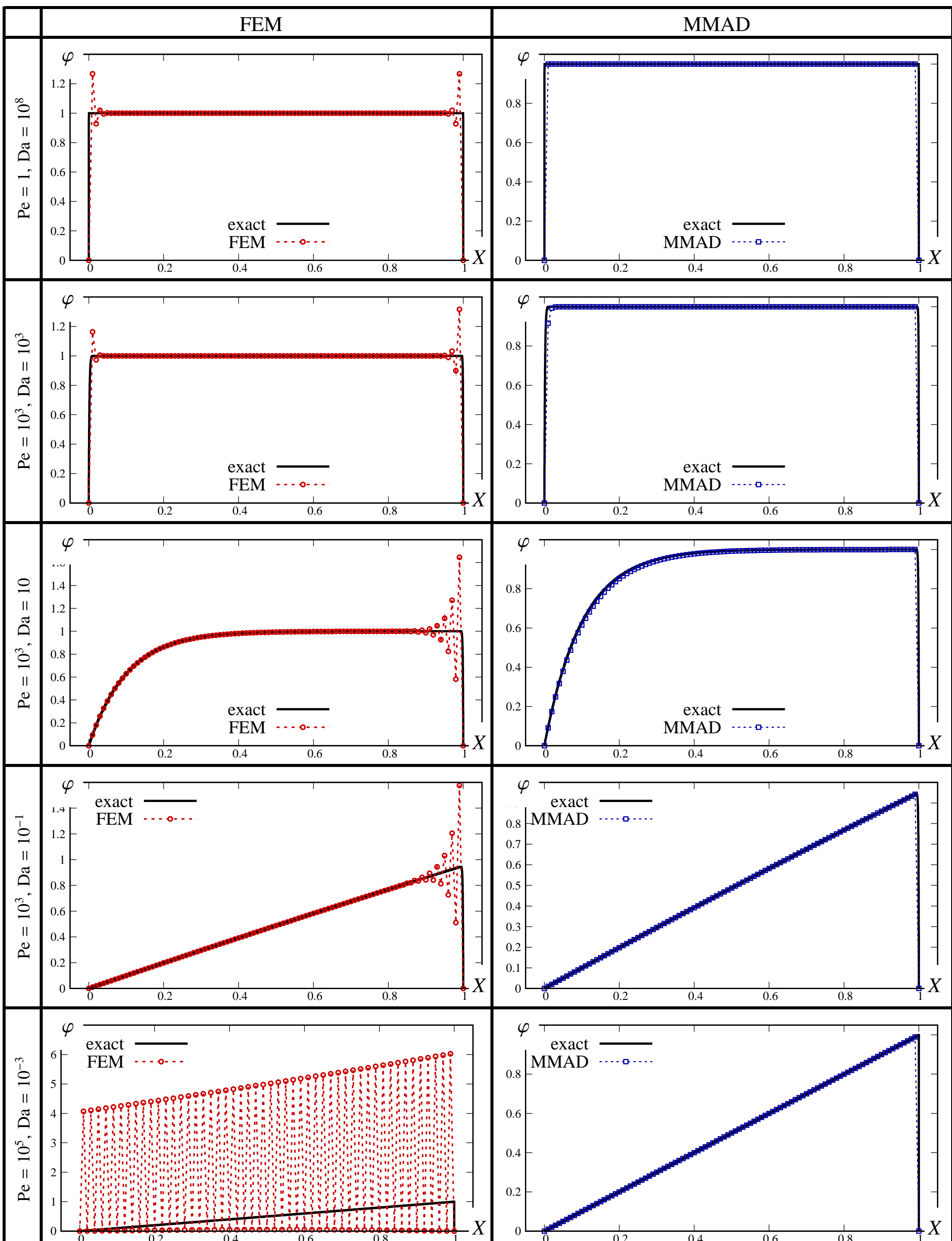

**Figure 2** Comparison of FEM and MMAD solution for a 1D reaction-convection-diffusion problem described in Example 1. The solid black line represents the exact solution and the dashed lines (with points on top) represent the solutions obtained from different numerical schemes. The characteristic element length is $h = 0.01$.

**Example 2** We consider a 2D reaction-convection-diffusion problem where the domain $\mathcal{B} := [0, 1] \times [0, 1]$ is subject to Dirichlet boundary conditions on two edges and Neumann boundary conditions on two other edges. The source term is $F = 0$, the element characteristic length is $h = 0.025$ and the flow is irrotational with angle 45°. This benchmark example is taken from (Brooks and Hughes 1982). Note though that, the results presented in (Brooks and Hughes 1982) do not include reaction.

Results obtained with standard finite element approximations and the current MMAD

formulation are presented in Figure 3.

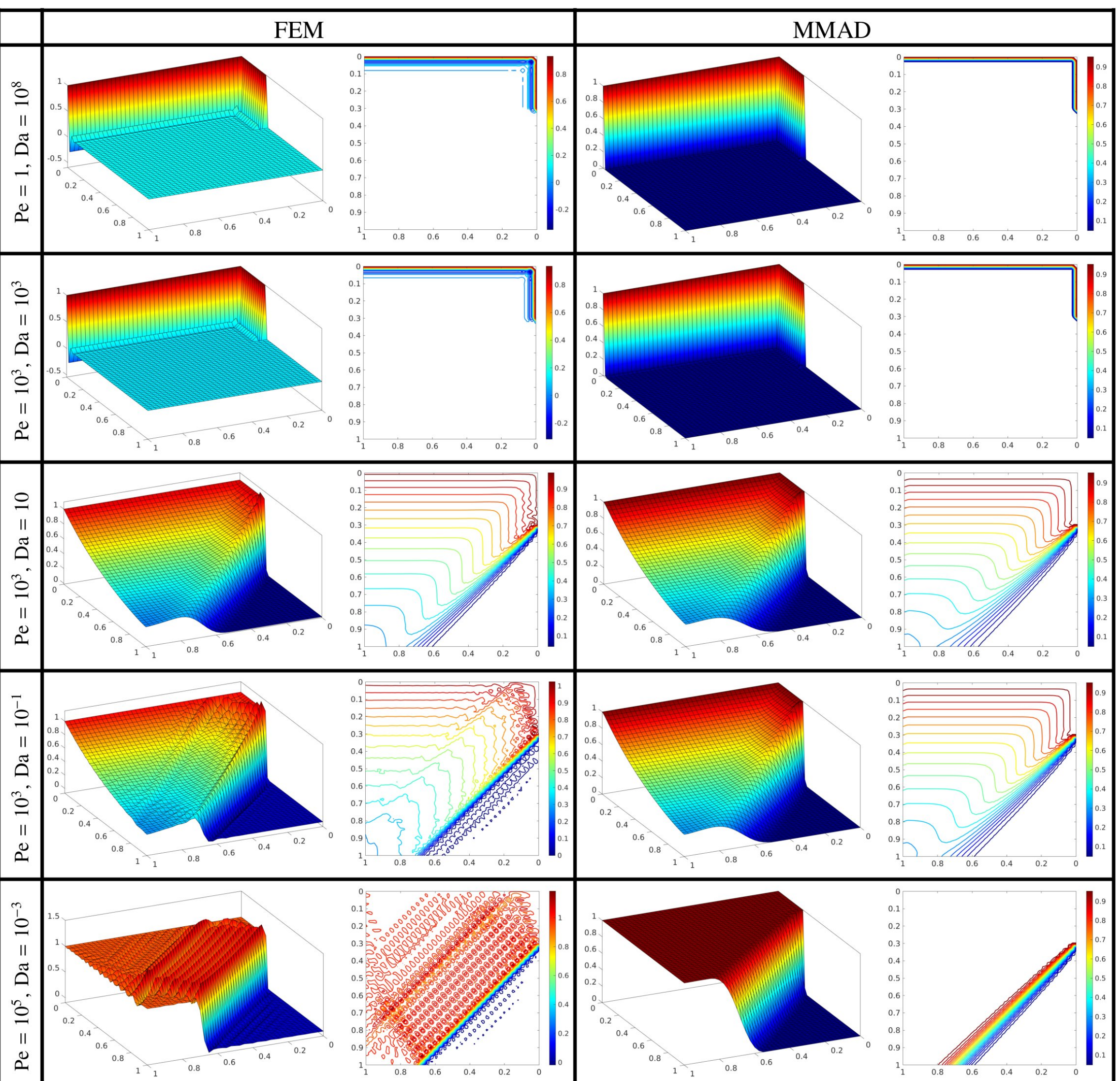

**Figure 3** Comparison of FEM and MMAD solution for a 2D reaction-convection-diffusion problem described in Example 2 taken from (Brooks and Hughes 1982). The characteristic element length is $h = 0.025$ and the flow direction is $\boldsymbol{u} = [\sqrt{2}/2, \sqrt{2}/2]$.

For the reaction-dominated case, the solution by FEM suffers from small oscillations close to the boundary. The same level of oscillations holds for the case with equal Damköhler and Peclet numbers. For the other cases with Peclet number greater than the Damköhler number, the oscillations close to the boundary disappear and they appear in the vicinity of the shock. Increasing the ratio of Peclet to Damköhler number leads to higher degrees of oscillation. The MMAD method though leads to an exact solution for the reaction-dominated case and the case with equal Damköhler and Peclet numbers.

For the remaining three cases where convection becomes more dominant, the MMAD solution is stable and relatively accurate. Note that the MMAD solution for these cases is not identical to the exact solution; that is, the shock in the exact solution is significantly sharper, occurring over the width of a single element, whereas the MMAD solution exhibits a more gradual transition.

**Remark 5** The exact analytical solution is shown only for the 1D case. For the 2D cases, instead of the exact solution to this problem, a reference solution obtained from a significantly refined mesh was used as a proxy for the exact solution. This solution exhibits very sharp gradients where the transition occurs within the width of a single element.

This reference solution is not included in the figures as the behavior is qualitatively reproduced by the MMAD method in the initial examples shown in Figures 3 and 4. Accordingly, we state that “the MMAD method leads to a highly accurate solution for the reaction-dominated case and the case with equal Damköhler and Péclet numbers," based on the excellent agreement with the highly resolved solution.

**Example 3** As suggested in (Brooks and Hughes 1982; Tezduyar and Park 1986), we consider a 2D reaction-convection-diffusion problem where the domain $\mathcal{B} := [0,1] \times [0,1]$ is subject to Dirichlet boundary conditions on all its edges. Similar to the previous example, the source term is $F = 0$, the element characteristic length is $h = 0.025$ and the flow is irrotational with angle 45°. Note that the results presented in (Brooks and Hughes 1982) do not include reaction.

Results obtained with standard FEM and the current MMAD formulation are presented in Figure 4. Similar to the previous example, for the reaction-dominated case and the case with equal Damköhler and Peclet numbers, the solution by FEM suffers from small oscillations close to the boundary. For the other cases with Pe $= 10^3$ and Da $= 10$, the oscillations exist both close to the boundary and in the vicinity of the shock. For the last two cases the FEM solutions are completely unstable. Similar observations to the previous example hold for the MMAD solutions.

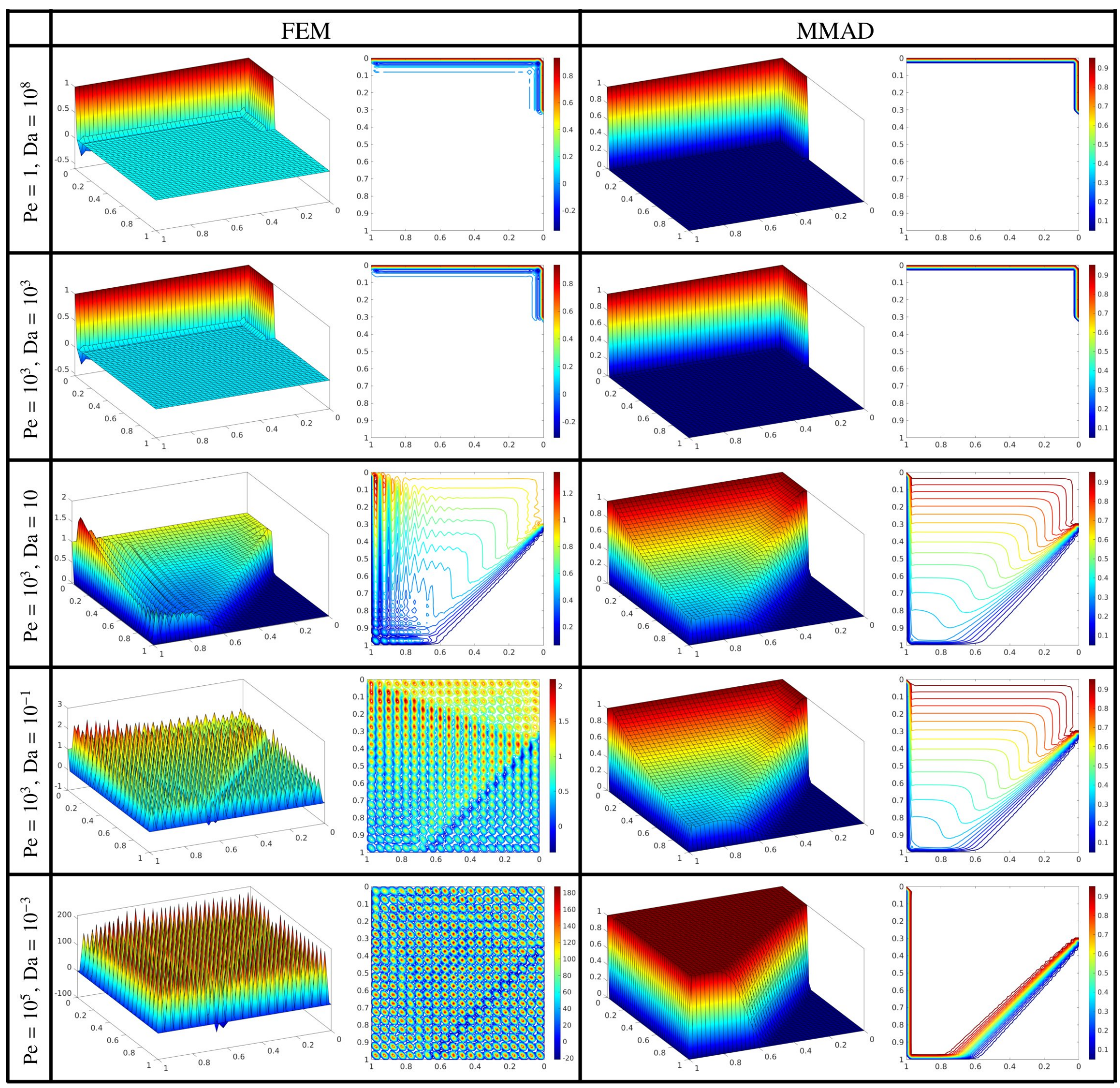

**Figure 4** Comparison of FEM and MMAD solution for a 2D reaction-convection-diffusion problem described in Example 3. The characteristic element length is $h = 0.025$ and the flow direction is $\boldsymbol{u} = [\sqrt{2}/2, \sqrt{2}/2]$.

**Example 4** As suggested in (Brooks and Hughes 1982; Tezduyar and Park 1986), we consider a 2D reaction-convection-diffusion problem with the domain $\mathcal{B} := [0,1] \times [0,1]$. As shown in Figure 1, all four edges are subject to Dirichlet boundary conditions $\varphi = 0$. The velocity components are defined as $u_1 = -x_2$ and $u_2 = x_1$ with $x_1$ and $x_2$ being the coordinates. Another condition is enforced on the line AB along which $\varphi$ is set equal to $\sin(2\pi x_2)$. The source term is $F = 0$, the element characteristic length is $h = 0.025$. The example is also referred to as the "Donut problem". Note that, the results presented in (Brooks and Hughes 1982) do not include reaction.

The results obtained with standard FEM and the MMAD method are presented in Figure 5. For the two first cases with $\mathrm{Da} = 10^6$, and $\mathrm{Da} = 1$ (respectively reaction-dominated and slightly convection-dominated), the hill is damped very early, nonetheless the FEM solution suffers from instabilities on the left side of the hill. For the third convection-dominated case with $\mathrm{Da} = 10^{-2}$ and $\mathrm{Pe} = 10^6$, the FEM solution renders severe instabilities all over the domain. Further decrease in the Damköhler number yields more dominance of convection leading to more stable solutions. The MMAD solution exhibits stable and accurate behavior regardless of the dominance of convection or reaction.

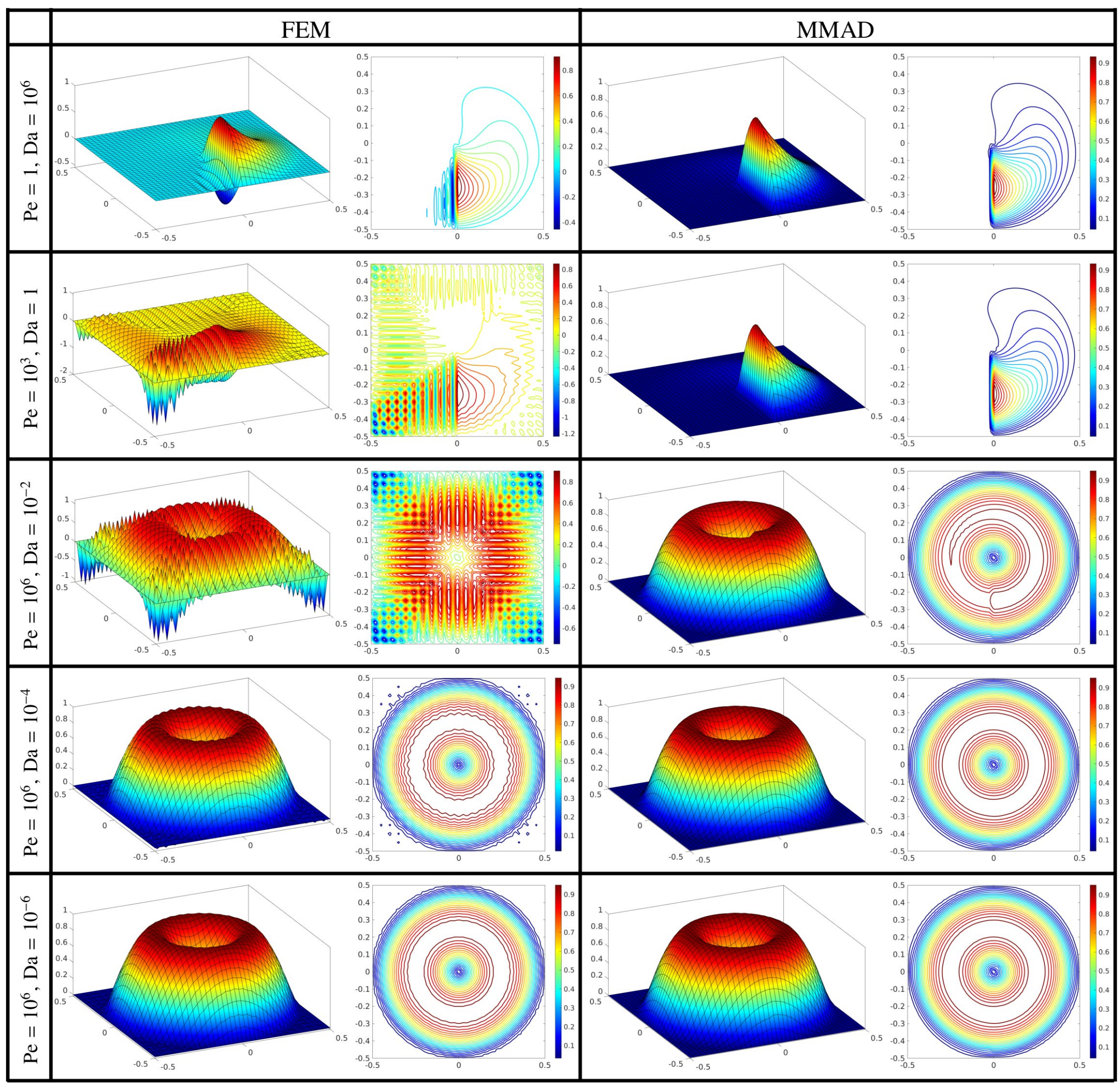

**Figure 5** Comparison of FEM and MMAD solution for a 2D reaction-convection-diffusion problem described in Example 4. The characteristic element length is $h = 0.025$ and the flow is rotational with the characteristic velocity $\boldsymbol{u} = [-x_2, x_1]$ with $x_1$ and $x_2$ being the coordinates.

**Example 5** As suggested in (Franca and Valentin 2000; Duan et al. 2012), we consider a 2D reaction-convection-diffusion problem where the domain $\mathcal{B} := [0, 1] \times [0, 1]$ is subject to Dirichlet boundary conditions on all its edges. The source term is $F = 0$, the element characteristic length is $h = 0.025$ and the flow is irrotational with velocity components $\boldsymbol{u} = [0.15, 0.1]$. Results obtained with FEM and MMAD method are presented in Figure 6. Observations similar to Example 3 can be drawn regarding the behavior of the FEM and MMAD solutions.

**Example 6** We consider a 2D reaction-convection-diffusion problem where the domain $\mathcal{B} := [0, 1] \times [0, 1]$ is subject to Dirichlet boundary conditions on all its edges. The source term here is $F = 1$, the element characteristic length is $h = 0.025$ and the flow is irrotational with angle $60°$. This benchmark example is taken from (Codina 1998). The values for the velocity, diffusion coefficient and reaction coefficient are identical to those considered in (Codina 1998). That is, the diffusion coefficient is $D = 10^{-4}$ m$^2$/s, the velocity and reaction coefficient are chosen accordingly to yield the given Damköhler and Peclet numbers.

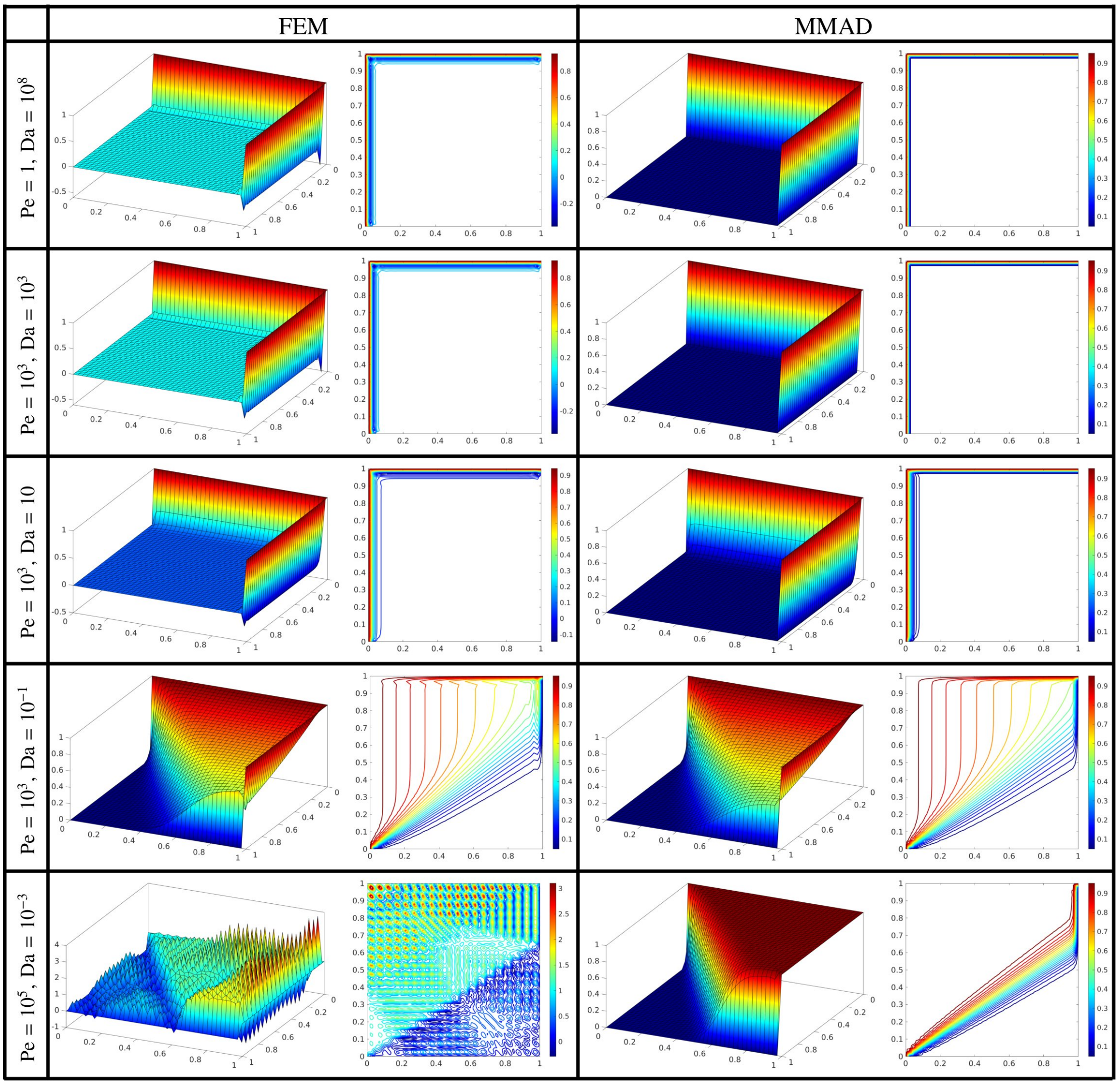

**Figure 6** Comparison of FEM and MMAD solution for a 2D reaction-convection-diffusion problem described in Example 5. The characteristic element length is $h = 0.025$ and the flow direction is $\boldsymbol{u} = [0.15, 0.1]$.

The results obtained with standard FEM and the MMAD method are presented in Figure 7. For the first and the last cases with high Peclet numbers, the FEM solution is completely oscillatory and unstable. Some oscillations close to the boundary are observed for the case with high Damköhler number. The MMAD solution provide stable and accurate solutions for all the three cases.

**Computational time comparison** Figure 8 compares the computational cost of the classical FEM and the MMAD method in both one-dimensional (1D) and two-dimensional (2D) settings. The figure shows the normalized computational time for each method, where the red bars represent FEM and the blue bars represent MMAD. MMAD is consistently more computationally expensive than FEM in both 1D and 2D. This is due to the introduction of the additional field $\boldsymbol{g}$, which increases the number of degrees of freedom (DOFs). In 1D, FEM has one DOF per node, while MMAD has two DOFs per node. In 2D, the difference is even more pronounced where FEM still has one DOF per node, but MMAD has three DOFs per node. This explains why the gap in computational cost between FEM and MMAD widens as the problem dimension increases.

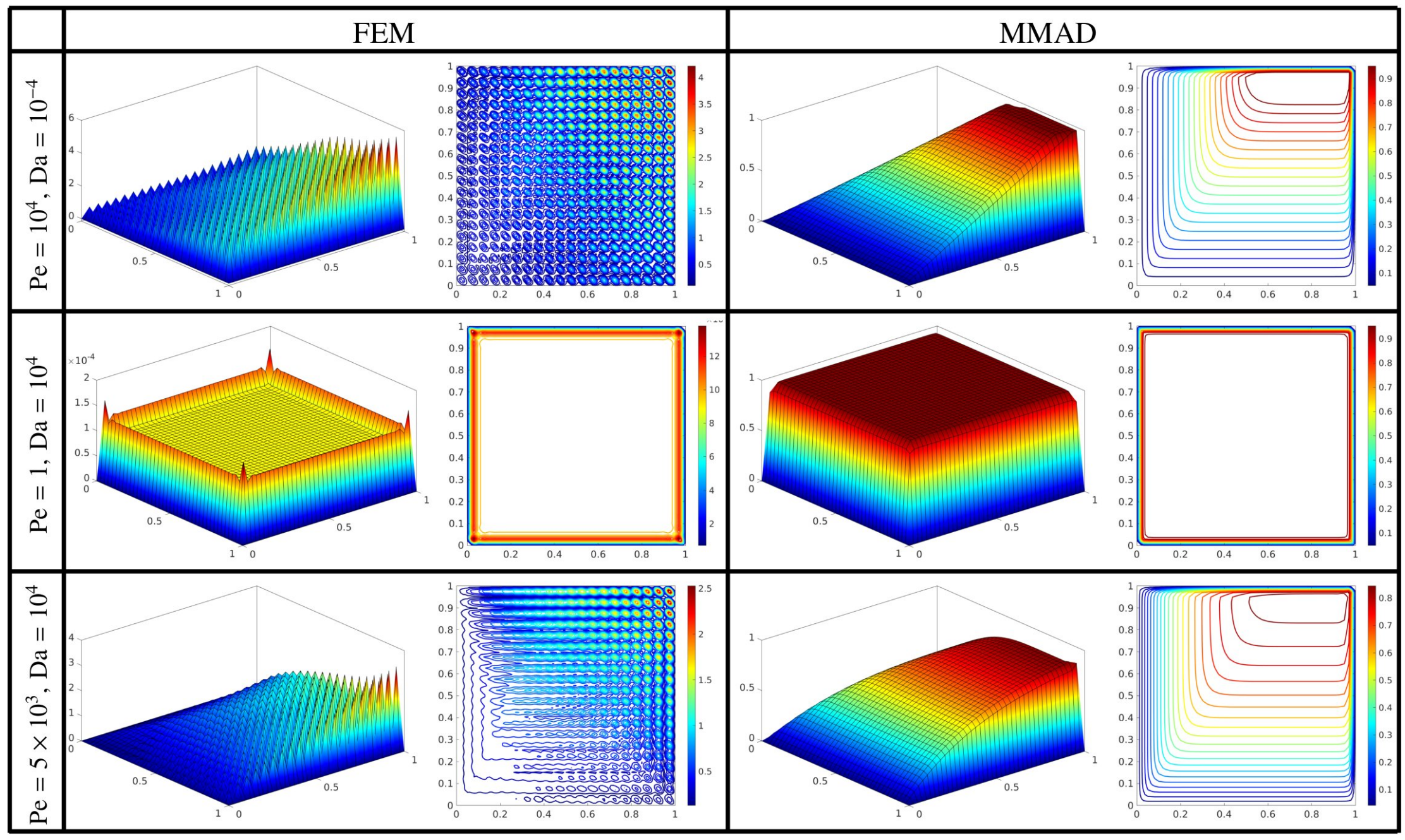

**Figure 7** Comparison of FEM and MMAD solution for a 2D reaction-convection-diffusion problem described in Example 6. The characteristic element length is $h = 0.025$ and the flow direction is $\boldsymbol{u} = [1/2, \sqrt{3}/2]$.

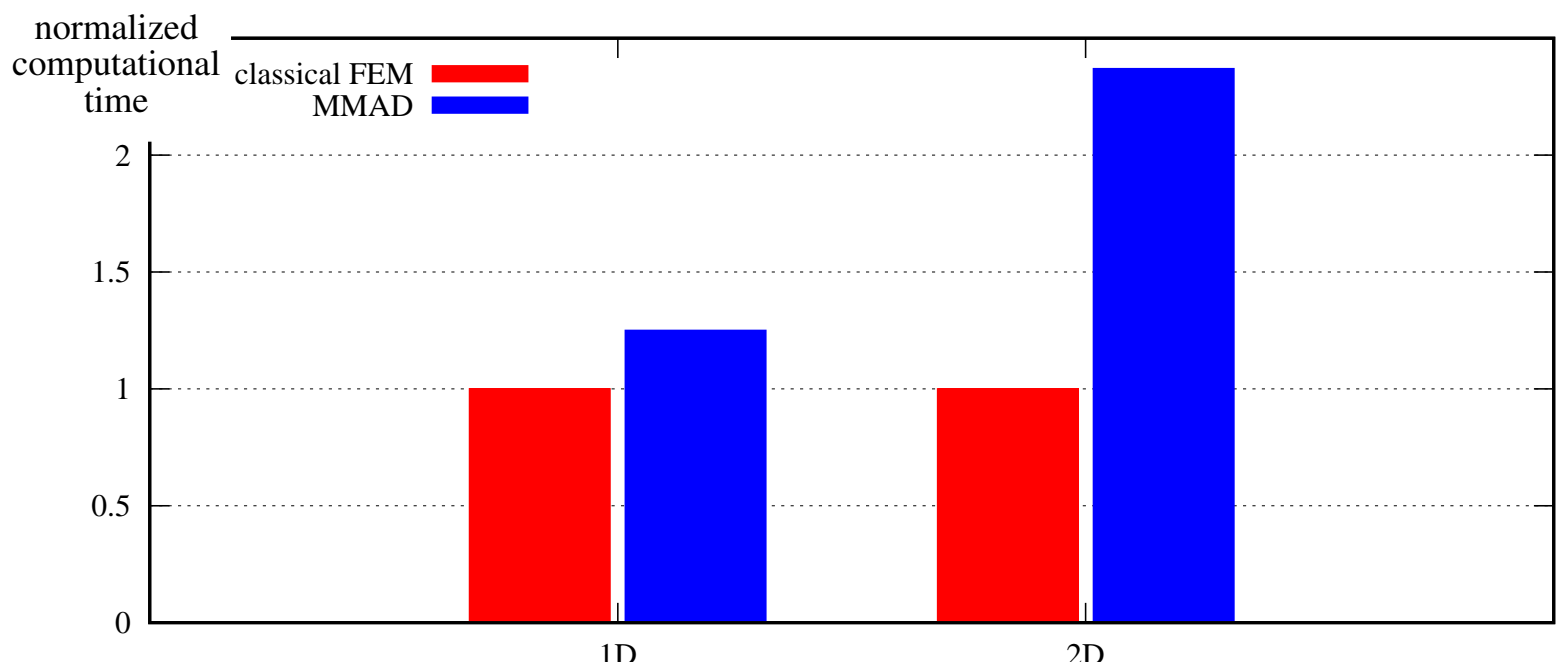

**Figure 8** Comparison of the computational time between the classical FEM and the MMAD method.

## 5 Summary and outlook

We have proposed a novel micromorphic-based artificial diffusion method to circumvent the instabilities commonly observed in standard Galerkin finite element approximations of reaction-convection-diffusion problems. The key idea in our approach is to introduce an auxiliary micromorphic-type variable, which enriches the formulation through the addition of terms involving the variable and its gradient. Sufficient conditions for well-posedness and convergence of the method have been presented. A number of examples comprising reaction- or convection-dominated situations, as well as combinations of two, illustrate the stability and accuracy of the MMAD approach. An important challenge in extending our proposed approach arises in the context of fully coupled flow problems where the velocity field is no longer prescribed but instead treated as an additional unknown. In such settings, the micromorphic coupling tensor $\boldsymbol{H}$, which is designed to align with the local flow direction, must be determined dynamically. Developing adaptive or robust strategies for defining $\boldsymbol{H}$ in such coupled problems is an important direction for future research.

**Authors' contributions** SF: Conceptualization, Methodology, Software, Formal Analysis, Visualization, Writing – Original Draft, Writing – Review  Editing. DR: Supervision, Validation, Mathematical Proof, Review  Editing. PS: Supervision, Project Administration, Review  Editing

**Funding** SF and PS gratefully acknowledge the support by the Deutsche Forschungsgemeinschaft (DFG, German Research Foundation) project number 460333672 – CRC 1540 Exploring Brain Mechanics (subproject C01).

**Competing interests** The authors declare that they have no competing interests.